\documentclass[preprint,12pt]{elsarticle}

\usepackage{graphicx}
\usepackage[fleqn]{amsmath}
\usepackage{booktabs}
\usepackage{float}
\usepackage{color}
\usepackage{soul} 
\usepackage{lineno}

\usepackage{lineno}

\let\oldequation\equation
\let\oldendequation\endequation

\renewenvironment{equation}
  {\linenomathNonumbers\oldequation}
  {\oldendequation\endlinenomath}

\journal{PLB}

\begin{document}

\begin{frontmatter}



\title{Neutron radius determination of $^{133}$Cs and its impact on the interpretation of CE$\nu$NS-CsI measurement}


\author[label1,label2]{Y. Huang}
\author[label3,label4]{S. Y. Xia}
\author[label3,label4]{Y. F. Li}
\author[label1]{X. L. Tu\corref{cor2}}
\cortext[cor2]{Corresponding author.}
\ead{tuxiaolin@impcas.ac.cn}
\author[label1]{J. T. Zhang}
\author[label1]{C. J. Shao}
\author[label1]{K. Yue}
\author[label1]{P. Ma}
\author[label5]{Y. F. Niu}
\author[label6]{Z. P. Li}
\author[label6]{Y. Kuang}
\author[label2]{X. Q. Liu}
\author[label2]{J. F. Han}
\author[label7]{P. Egelhof}
\author[label7]{Yu.  A. Litvinov}
\author[label1]{M. Wang}
\author[label1]{Y. H. Zhang}
\author[label1]{X. H. Zhou}
\author[label1]{Z. Y. Sun}

\address[label1]{Institute of Modern Physics, Chinese Academy of Sciences, Lanzhou 730000,China}
\address[label2]{Key Laboratory of Radiation Physics and Technology of the Ministry of Education, Institute of Nuclear Science and Technology, Sichuan University, Chengdu 610064, China}
\address[label3]{Institute of High Energy Physics, Chinese Academy of Sciences, Beijing 100049, China}
\address[label4]{School of Physical Sciences, University of Chinese Academy of Sciences, Beijing 100049, China}
\address[label5]{School of Nuclear Science and Technology, Lanzhou University, Lanzhou 730000, China}
\address[label6]{School of Physical Science and Technology, Southwest University, Chongqing 400715, China}
\address[label7]{GSI Helmholtzzentrum f\"ur Schwerionenforschung GmbH, D-64291 Darmstadt, Germany}

\begin{abstract}
Proton-$^{133}$Cs elastic scattering at low momentum transfer is performed using an in-ring reaction technique at the Cooler Storage Ring at the Heavy Ion Research Facility in Lanzhou.
Recoil protons from the elastic collisions between the internal H$_2$-gas target and the circulating $^{133}$Cs ions at 199.4 MeV/u are detected by a silicon-strip detector. The matter radius of $^{133}$Cs is deduced by describing the measured differential cross sections using the Glauber model.
Employing the adopted proton distribution radius, a point-neutron radius of 4.86(21) fm for $^{133}$Cs is obtained.
With the newly determined neutron radius, the weak mixing angle sin$^2 \theta_W$ is independently extracted to be 0.227(28) by fitting the coherent elastic neutrino-nucleus scattering data. 
Our work limits the sin$^2 \theta_W$ value in a range smaller than the ones proposed by the previous independent approaches, and would play an important role in searching new physics via the high precision CE$\nu$NS-CsI cross section data in the near future.
\end{abstract}


\end{frontmatter}







The weak mixing angle, sin$^2\theta_W$, is a fundamental parameter in the $SU(2)_{\rm L}\times U(1)_{\rm Y}$ electroweak theory of the Standard Model (SM)~\cite{Suekane21}.
Any deviation from the expected sin$^2\theta_W$ value in SM may serve as an indicative signature of new physics.
Historically, the masses of top quark and Higgs boson were successfully predicted by the higher order diagram calculations with the measured sin$^2\theta_W$~\cite{Suekane21}.

To precisely constrain the sin$^2\theta_W$ value, various methods are developed to measure the dependence of sin$^2\theta_W$ on the transferred momentum~\cite{Kumar13,ALEPH06}. 
Very recently, a high precise of about 0.1$\%$ for the sin$^2\theta_W$ determination was achieved by the CMS experiment~\cite{CMS}.
At very low momentum transfer, for instance, one uses the atomic parity nonconservation (PNC) measurements of $^{133}$Cs~\cite{Dzuba12,Wood97}. However, as pointed out in Ref.~\cite{Cadeddu23}, the sin$^2\theta_W$ determination via the PNC depends on many theoretical corrections. There is an obvious difference for the PNC amplitude correction associated with neutron skin effects~\cite{Sahoo21}. The effects of the unknown neutron distribution radius of $^{133}$Cs on the PNC were addressed about 25 years ago~\cite{Pollock99}.

Recently the coherent elastic neutrino-nucleus scattering (CE$\nu$NS) was observed using a CsI[Na] detector at the Oak Ridge National Laboratory~\cite{Akimov17}. 
The CE$\nu$NS measurement allows one to investigate many physics, for instance, neutrino nonstandard interactions~\cite{Akimov221}, dark matter~\cite{Akimov23}, and light vector $Z^{'}$ mediator~\cite{Cadeddu2021jh}, see Refs.~\cite{Cadeddu23,Barbeau23} for details.
Specially, the CE$\nu$NS experiment also provides a clear method to constrain sin$^2\theta_W$ at low momentum transfer~\cite{Akimov221}.
Different from other experimental methods~\cite{Kumar13,ALEPH06}, the sin$^2\theta_W$ deviation between the CE$\nu$NS experiment and the SM prediction could give a hint on new $\nu$-nucleon interactions~\cite{Huang19}.
Although majority of observables were experimentally measured, there was still a lack of precision neutron radii of $^{133}$Cs and $^{127}$I~\cite{Akimov221,Cadeddu2021jh,Sierra19} in interpreting the CE$\nu$NS-CsI data. 
As underscored by the COHERENT collaboration~\cite{Akimov221}, the uncertainty in the CE$\nu$NS-CsI cross section calculations is dominated by the neutron distributions of $^{133}$Cs and $^{127}$I. 
The influence of the neutron radii on the CE$\nu$NS-CsI cross section data interpretation was also addressed by other independent investigations~\cite{Sierra19}.
In a word, inaccurate or biased treatment of the neutron radii of $^{133}$Cs and $^{127}$I would lead to the misidentification of possible signals of new physics~\cite{Sierra19}.
Usually, sin$^2\theta_W$ can be deduced through a fit to the CE$\nu$NS-CsI data by fixing neutron radii from theoretical predictions of various models~\cite{Akimov221,Papoulias20,Khan19,Cadeddu20}.
Alternatively, the neutron radii can be deduced by assuming a fixed sin$^2\theta_W$ value~\cite{Cadeddu18}. 
Up to date, the reported (average) neutron radii of $^{133}$Cs spread from 4.6 fm through 6.6 fm~\cite{Huang19,Papoulias20,Khan19,Cadeddu18,Cadeddu201,Cadeddu21,Coloma20,Cadeddu19,Romeri23,Papoulias20_1,Corona23}.

The accurate determination of the neutron radius for $^{133}$Cs has garnered significant attention across the atomic, nuclear, and particle physics communities owing to its significance in fundamental researches.
However, it is challenging to experimentally measure the neutron radius of $^{133}$Cs in normal kinematics, due to the low melting point of 28$^{\circ}$C and spontaneous ignition in air. 
In this Letter, we introduce an innovative approach to determine the neutron radius of $^{133}$Cs by measuring the proton elastic scattering at low momentum transfer using an in-ring reaction technique and inverse kinematics. We elucidate the impact of this determination on the sin$^2 \theta_W$ extraction from the recent CE$\nu$NS-CsI data.

\begin{figure}[htbp]
\includegraphics*[width=9cm]{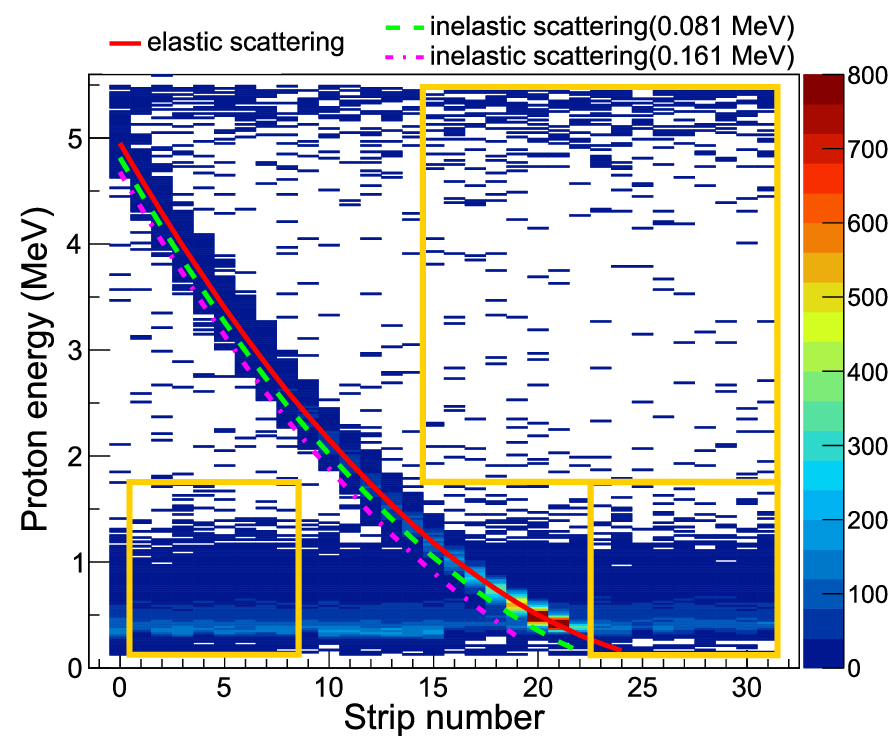}
\caption{Scatter plot of the recoil proton energy versus the strip number of DSSD.
The solid (red), dashed (green), and dash-dotted (pink) lines denote the calculated proton energies for elastic and two inelastic scattering channels, respectively.
For more details see text.}
\label{fig1}
\end{figure}

The novel in-ring reaction experiment was carried out at the experimental Cooler Storage Ring (CSRe) of the Heavy Ion Research Facility in Lanzhou (HIRFL)~\cite{Xia02}.
Such kinds of experiments are especially suited for the small-angle differential cross section measurements at low momentum transfer~\cite{Liu20,Zhang23}.
CSRe, which is equipped with the electron cooler~\cite{Mao16} and the internal H$_2$-gas-jet target~\cite{Shao13}, was operated for an in-ring reaction experiment at a magnetic rigidity of about 5.205 Tm.
The $^{133}$Cs$^{27+}$ beam with an energy of 204 MeV/u from the main storage ring (CSRm) was stripped off all bound electrons utilizing an aluminum foil with a thickness of 0.21 mm at the radioactive ion beam line (RIBLL2).
Then the $^{133}$Cs$^{55+}$ ions with an energy of about 199.4 MeV/u were transported through RIBLL2, and injected into CSRe.
The stored $^{133}$Cs$^{55+}$ ions in CSRe interacted repeatedly with the H$_2$-gas target of about 10$^{12}$ atoms/cm$^2$ thickness~\cite{Shao13}.
The electron cooling at CSRe was operated to compensate for the energy loss of the ions caused by the collisions with the gas target and residual gas.
The recoil protons from the $p$-$^{133}$Cs elastic scattering within angular range from about 85$^{\circ}$ to 90$^{\circ}$ in the laboratory system were measured by a double-sided silicon-strip detector (DSSD) with a typical energy resolution of better than 1$\%$.
The employed 1000 $\mu$m thick DSSD had an active area of 64 $\times$ 64 mm$^2$ and was segmented into 32 $\times$ 32 strips.
The proton energy and detection efficiency were calibrated by radioactive sources~\cite{Zhang23}.
Figure~\ref{fig1} illustrates the scatter plot of the recoil proton energy versus the strip number of DSSD.
Further experimental details were described in our previous works~\cite{Zhang23,Yue19,Zhang19}.

\begin{figure}[htbp]
\includegraphics*[width=9cm]{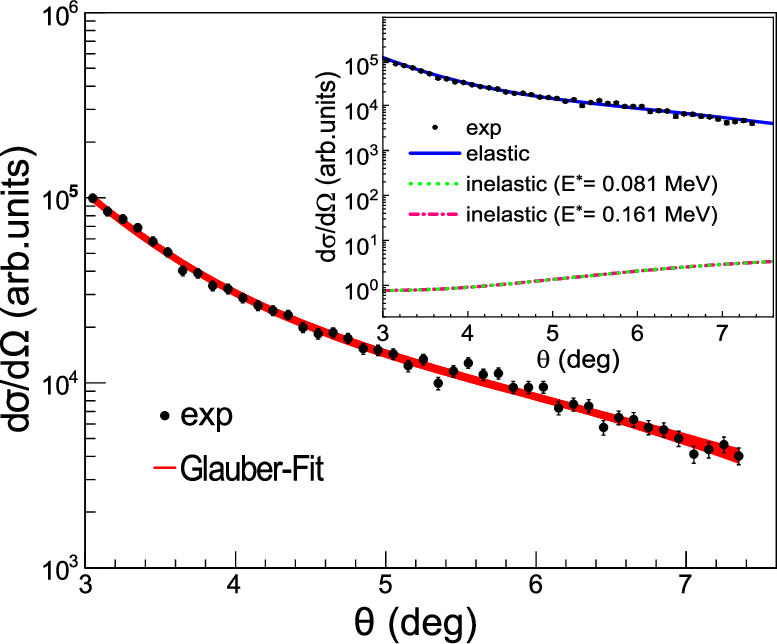}
\caption{The measured $\frac{d\sigma}{d\Omega}(\theta)$ for $p$-$^{133}$Cs elastic scattering and the 2$\sigma$ distribution of fit curves (red).
The insert shows the calculated elastic and inelastic differential cross sections using FRESCO~\cite{Thompson88}. 
The trend of the measured $\frac{d\sigma}{d\Omega}(\theta)$ values agree well with the FRESCO calculations.
Conveniently $\frac{d\sigma}{d\Omega}(\theta)$ are normalized to the FRESCO calculations.
 }
\label{fig2}
\end{figure}

Given the fact that the flight paths and energies of recoil protons are hardly altered by secondary collisions with the thin gas target, the relative small-angle differential cross sections $\frac{d\sigma}{d\Omega}(\theta)$ of $p$-$^{133}$Cs elastic scattering are determined via~\cite{Zhang23}
\begin{equation}
\frac{d\sigma}{d\Omega}(\theta)=\frac{1}{\sin\theta}\left(\frac{\Delta N_{\rm all}}{\Delta \theta}-\frac{\Delta N_{\rm bg}}{\Delta \theta}\right),
\end{equation}
where $\Delta N_{\rm all}$ is the number of all measured events in the scattering angle interval $\Delta \theta$ in the center-of-mass (c.m.) frame, and $\Delta N_{\rm bg}$ is the corresponding background estimated by the measured events in the rectangles enclosed by the yellow solid lines, see Fig.~\ref{fig1}.
The scattering angles $\theta$ were determined by the proton kinematic energies $K_{\rm lab}$ via the relation of $2m_{p}K_{\rm lab}=2p^2(1-\cos\theta)$,
with $m_{p}$ and $p$ being the proton rest mass and the c.m. momentum, respectively.
The uncertainties of $\theta$ are smaller than the used $\Delta \theta$ value of 0.1$^{\circ}$.
To reduce the effects of solid angle and detection efficiency~\cite{Zhang23}, only single coincidence events between $X$ and $Y$ strips within the energy range of about 0.6 MeV through 3.6 MeV were considered.
As shown in Fig.~\ref{fig1}, the events of elastic scattering are mixed, to some extent, with those of inelastic scattering associated with the low-lying 0.08- and 0.16-MeV excited states~\cite{Khazov11}. 
However, experiments have already shown that the inelastic scattering cross sections are very tiny in the small angular region, compared to the elastic scattering cross sections~\cite{McDaniels86, Kailas84}.
For the $^{133}$Cs case, according to the FRESCO calculations~\cite{Thompson88} with the phenomenological optical model~\cite{Koning03}, the inelastic scattering cross sections are also several orders of magnitude smaller than those of the elastic scattering in the measured small angular range, see the insert of Fig.~\ref{fig2}.
Therefore, the contribution of the inelastic scattering can be safely ignored, compared to the several percent relative uncertainties of experimental differential cross sections.
The measured small-angle $\frac{d\sigma}{d\Omega}(\theta)$ values are shown in Fig.~\ref{fig2}.

It is well known that the small-angle elastic $\frac{d\sigma}{d\Omega}(\theta)$ distributions are sensitive to matter distribution radius~\cite{Liu20,Zhang23,Alkhazov97,Korolev18,Schmid23,Huang232}.
Especially, the reaction mechanism is relatively simple at small angles and thus correction terms of reaction models are negligible compared to the cases of large angle scattering~\cite{Alkhazov78,Kudryavtsev89}.
In addition, the model-dependent errors related to the hadronic probes~\cite{Thiel19} can also be effectively reduced through calibrating the input parameters of reaction models using the well-known nuclear radii~\cite{Zhang23,Terashima08,Huang23,Alkhazov02}.
Therefore, many experimental setups at various facilities are developed to determine the matter radii by measuring the small-angle $\frac{d\sigma}{d\Omega}(\theta)$ of $p$-nucleus elastic scattering~\cite{Liu20,Zhang23,Yue19,Alkhazov97,Korolev18,Schmid23}.
In the present work, a well established procedure~\cite{Alkhazov78,Alkhazov02,Huang231} based on the Glauber multiple-scattering theory~\cite{Glauber59} is employed to extract the matter radius of $^{133}$Cs through describing the measured $\frac{d\sigma}{d\Omega}(\theta)$. 
The $\frac{d\sigma}{d\Omega}(\theta)$ values are expressed in the Glauber model as a function of the matter density distribution $\rho(r)$ and the proton-nucleon scattering amplitude $f_{pi}(q)$ with $i=n$ or $p$, see Ref.~\cite{Huang231} for details.
To reduce the model-dependent errors of matter radius, the scattering amplitude parameters were calibrated at 200 MeV~\cite{Huang23} to be $\sigma_{pp}=$ 1.788(20) fm$^2$, $\sigma_{pn}=$ 3.099(27) fm$^2$, $\alpha_{pp}=$ 0.893(17), $\alpha_{pn}=$ 0.325(23), and $\beta_{pp}=\beta_{pn}=$ 0.528(41) fm$^2$, which are adopted here to calculate the $f_{pi}(q)$.
These values have been adopted to fit the differential cross sections of $p$-$^{16}$O elastic scattering at 200 MeV and reproduce the well-known matter radius of $^{16}$O~\cite{Huang23}.

In the radius fitting procedure, the $\rho(r)$ is described by the two-parameter Fermi model as
\begin{equation}
\rho(r)=\frac{\rho(0)}{1+{\rm exp}\left[(r-R)/{a}\right]},
\end{equation} 
with $\rho(0)$, $R$, and $a$ being the density normalization factor, half-density radius, and diffuseness parameter, respectively. 
Although the $R$ and $a$ values cannot be simultaneously constrained by $\frac{d\sigma}{d\Omega}(\theta)$ in small angular range~\cite{Zhang23}, the matter radius is almost independent of $a$ in the range of 0.50 fm through 0.55 fm for the medium-heavy nuclei~\cite{Huang231}.
Thus, following the methods in~\cite{Zhang23,Huang231,Tanaka20,Suzuki95}, we set $a=0.53(3)$ fm, which was deduced from the neighbouring $^{116,124}$Sn and $^{208}$Pb nuclei~\cite{Huang231}.
Additionally, a free cross section normalization factor $L_0$ is introduced as in Refs.~\cite{Zhang23,Huang232,Huang231} to reduce the radius uncertainty from the absolute $\frac{d\sigma}{d\Omega}(\theta)$ normalization. 
Subsequently, $R$ and $L_0$ are freely adjusted to fit the experimental $\frac{d\sigma}{d\Omega}(\theta)$ with the Glauber model.

As shown in Fig.~\ref{fig2}, the measured $\frac{d\sigma}{d\Omega}(\theta)$ are well described with the Glauber model by adjusting $R$ and $L_0$. 
With the obtained $R$ and fixed $a$, a root-mean-square (rms) point-matter radius $R_{\rm pm}$ for $^{133}$Cs is determined to be
\begin{equation}
R_{\rm pm} = \left(\frac{\int\rho(r)r^4dr}{\int\rho(r)r^2dr}\right)^{\frac{1}{2}}=4.811\pm0.127\;{\rm fm}\,,
\end{equation}
where uncertainties from statistics, input parameters, and Glauber model are about 0.12 fm, 0.03 fm, and 0.03 fm, respectively.
The radius uncertainties caused by statistics and input parameters are estimated by using the randomly sampled experimental $\frac{d\sigma}{d\Omega}(\theta)$ and input parameters within 2$\sigma$ band~\cite{Huang231}, respectively.
The model-dependent error at 200 MeV is estimated by comparing the well-known proton radii with the matter radii of $^{12}$C, $^{16}$O, and $^{28}$Si determined with the similar method, where similar proton and matter radii are expected for the $N=Z$ nuclei.
To check the effects of background, only recoil protons with energies $\textgreater$ 1 MeV were analyzed, and a consistent radius of 4.825 fm is obtained. 
Details and reliability considerations about radius determinations can be found in Refs.~\cite{Zhang23,Huang231}.

With the obtained $R_{\rm pm}$, a point-neutron distribution radius $R_{\rm pn}$ of $^{133}$Cs is determined to be 
\begin{equation}
R_{\rm pn}=\sqrt{\frac{A}{N}R_{\rm pm}^2-\frac{Z}{N}R_{\rm pp}^2}=4.86\pm0.21\; {\rm fm}, 
\end{equation}
where $N$, $Z$, and $A$ are the neutron, proton, and mass number, respectively.
The adopted point-proton radius $R_{\rm pp}$ of 4.740(5) fm for $^{133}$Cs is deduced from charge radius~\cite{Zhang23,Angeli13}.

We extract the neutron skin of $^{133}$Cs to be $R_{\rm pn}$$-$$R_{\rm pp}=0.12(21)$ fm. Meanwhile, using the linear relationships, established by various effective interactions~\cite{2021Cadeddu}, we also derive the neutron skin to be 0.14(3) fm from the Parity-violating electron scattering data of $^{48}$Ca and  $^{208}$Pb~\cite{Adhikari21,Adhikari22}. The two values are consistent with each other, and agree to the value of 0.13(4) fm calculated by the empirical linear relationship from the antiprotonic atom experiment~\cite{Trzcinska01}. The latter was adopted to estimate the PNC amplitude correction associated with the neutron skin effects~\cite{Derevianko01}.

Compared to the SM prediction, any deviation of the effective sin$^2 \theta_W$ determined by the $\rm CE\nu NS$-CsI data would be an indicative signature of new physics associated with the $\nu$ interactions.
Now we discuss the impact of the neutron radius on the sin$^2 \theta_W$ determination employing the $\rm CE\nu NS$-CsI cross section data~\cite{Akimov221}. 
For the description of the $\rm CE\nu NS$-CsI data in SM,
sin$^2\theta_W$ and the neutron distribution form factor $F_{n}(q^{2})$ are involved, see Eq.~(1) in Ref.~\cite{Corona23} for details. 
The folded-neutron radii $R_{\rm fn}$(Cs) and $R_{\rm fn}$(I) for $^{133}$Cs and $^{127}$I are indispensable to determine the corresponding $F_{n}(q^{2})$. 
Previous analyses indicated that different parameterizations of nuclear form factors would not lead to different sin$^2 \theta_W$~\cite{Huang19,Cadeddu2021jh,Corona23}.
In this work, the Helm form factor is adopted to calculate the expected $\rm CE\nu NS$-CsI signal event number, which is related to $R_{\rm fn}$ through the diffraction radius $R_{0}$ defined as~\cite{Corona23}
\begin{equation}
R_{0}(\rm K)= \sqrt{{5}/{3}[ \textit R_{\rm fn}(\rm K)^{2}-3s^{2}]},
\end{equation}
where $s = 0.9$ fm is the surface thickness and $\rm K=\rm I$ or Cs.
The $R_{\rm fn}$ can be directly determined from the point-neutron radius via $\left[ R_{\rm pn}^2+(0.864~\rm fm )^2\right]^{1/2}$~\cite{Cadeddu20}.

\begin{figure}[htbp]
\includegraphics*[width=9cm]{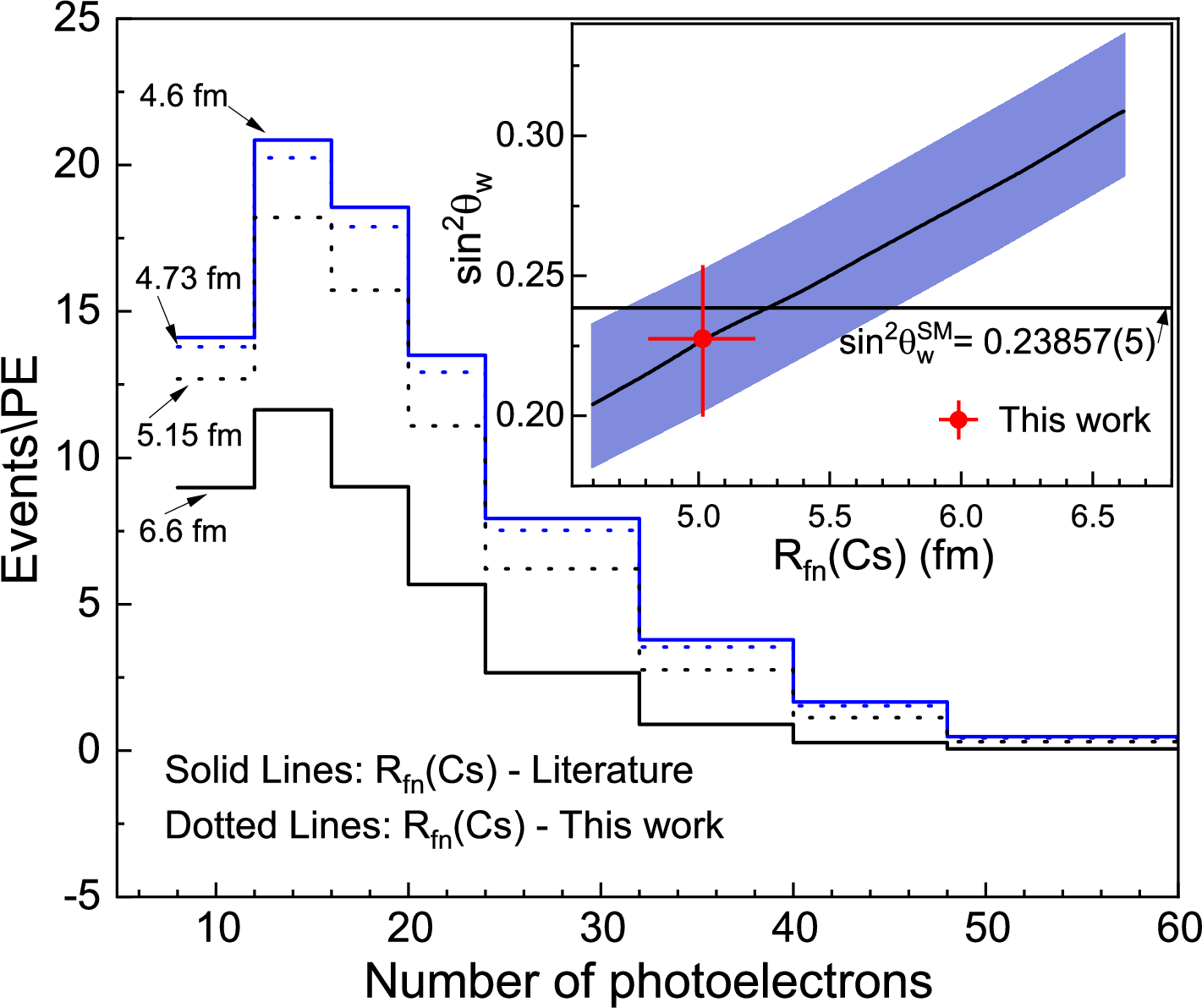}
\caption{The expected $\rm CE\nu NS$-CsI signal event number as a function of the photoelectron number calculated by using different neutron radii of $^{133}$Cs.
The blue and black solid lines are obtained by using the reported (average) neutron radius values of 4.6 fm~\cite{Khan19} and 6.6 fm~\cite{Corona23} for $^{133}$Cs, respectively, see Fig.~\ref{fig4}(b).
The blue and black dotted lines are calculated with the presently deduced radii $R^{\rm exp}_{\rm fn}(\mathrm{Cs})$ of $4.94\pm0.21$ fm, respectively.
The insert shows a correlation between $R_{\rm fn}$(Cs) and sin$^2 \theta_W$, where the blue area is the uncertainty caused by the CE$\nu$NS-CsI data~\cite{Akimov221}.
The red symbol shows the obtained $R_{\rm fn}^{\rm fit}$(Cs) and sin$^2 \theta_W$ by Eq.~(6), details see text. }
\label{fig3}
\end{figure}

Figure~\ref{fig3} depicts the $\rm CE\nu NS$-CsI signal event number as a function of the photoelectron number expected with different $R_{\rm fn}$(Cs).
In this work, the radius $R_{\rm fn}$(I) is determined using the $R_{\rm fn}$(Cs) value via $R_{\rm fn}$(Cs)$-$$R_{\rm fp}$(Cs)+$R_{\rm fp}$(I) considering that $^{133}$Cs and $^{127}$I have almost the same neutron skin thicknesses because of similar $(N-Z)/A$ values~\cite{Trzcinska01}, where the folded-proton radius $R_{\rm fp}$(K) is determined by the well-known charge radius~\cite{Cadeddu20,Angeli13}.
It is evident that the expected event spectrum is significantly affected by the adopted neutron radius of $^{133}$Cs, which is similar as addressed in Ref.~\cite{Sierra19}.   
As indicated in the inset of Fig.~\ref{fig3}, there exists a strong correlation between the values of $R_{\rm fn}$(Cs) and sin$^2 \theta_W$ in the $\rm CE\nu NS$-CsI data analysis.
An incorrect neutron radius can thus introduce significant shift in the estimation of sin$^2 \theta_W$.

Previously, a method was used combining the CE$\nu$NS and PNC data to precisely constrain sin$^2 \theta_W$~\cite{Corona23}.
However, possible new physics in either the CE$\nu$NS or PNC process may be ignored in such analysis due to the assumption that the two processes give the same sin$^2 \theta_W$ value. 
In addition, the PNC amplitude corrections depend on theory~\cite{Cadeddu23}, which may result in different values of sin$^2 \theta_W$~\cite{Corona23}.

\begin{figure}[htbp]
\includegraphics*[width=9cm]{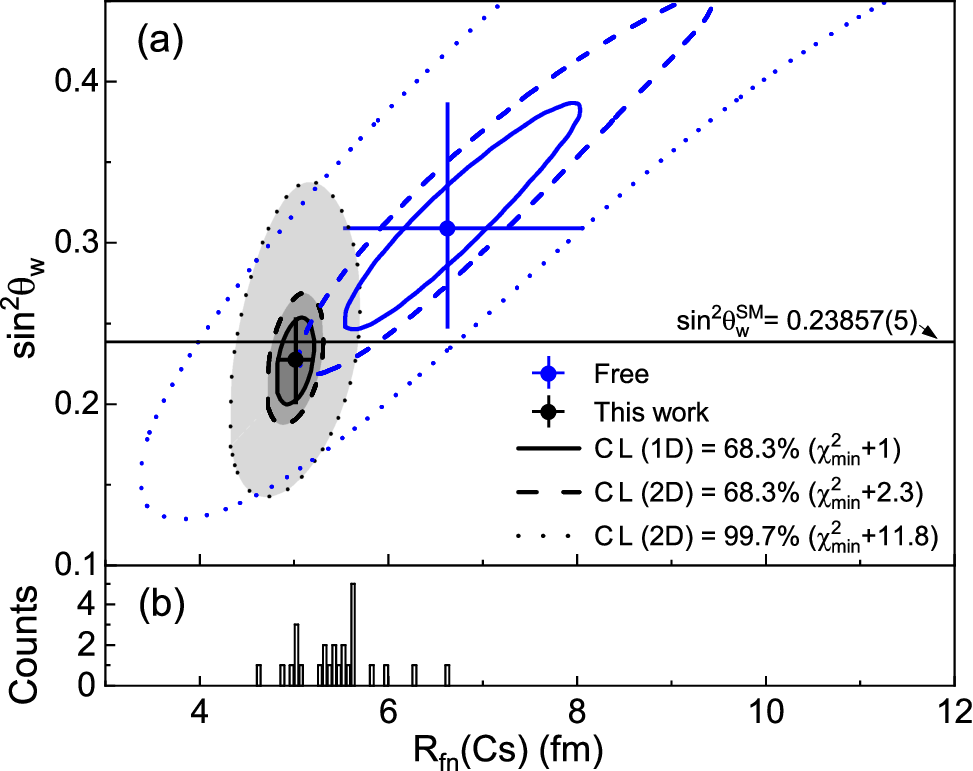}
\caption{(a) The $\chi^{2}_{\mathrm{all}}$ contours in the plane of $R_{\rm fn}$ versus sin$^2 \theta_W$. 
The blue curves and point represent results when both $R_{\rm fn}$ and sin$^2 \theta_W$ are free variables in the CE$\nu$NS-CsI data fitting.  
The black curves and point add the constraint imposed by the presently deduced radius.
(b) The distribution of reported neutron radii of $^{133}$Cs~\cite{Huang19,Papoulias20,Khan19,Cadeddu18,Cadeddu201,Cadeddu21,Coloma20,Cadeddu19,Romeri23,Papoulias20_1,Corona23} deduced from the CE$\nu$NS-CsI data~\cite{Akimov17,Akimov221}.  
}
\label{fig4}
\end{figure}

To avoid any improper input of neutron radius and sin$^2 \theta_W$, we performed an independent two-dimensional (2D) fit as Refs.~\cite{Huang19,Corona23} using both $R_{\rm fn}$ and sin$^2 \theta_W$ as free adjustable parameters. 
The sin$^2 \theta_W$ value from the best fit is shown in Fig.~\ref{fig4} as the blue symbol. 
The blue curves represent the distributions of the sin$^2 \theta_W$ values obtained from fitting procedures under different confidence levels (CL).  
Because of the strong correlation of the two free parameters, the independent 2D fit can not yield a well-constrained sin$^2 \theta_W$ value, nor the realistic neutron radius, as shown in Fig.~\ref{fig4}.  
Therefore accurate neutron radius is essential to deduce the sin$^2 \theta_W$ value and then to search for new physics beyond SM.

To show the impact of our newly determined neutron radius on the sin$^2 \theta_W$ value, the expected $\rm CE\nu NS$ events are fitted to the experimental events by adjusting sin$^2 \theta_W$ and $R^{\rm fit}_{\rm fn}(\mathrm{Cs})$ via the chi-square function defined here as 
\begin{equation}
\chi^{2}_{\mathrm{all}} = \chi^{2}_{\mathrm{CsI}}\left(R_{\rm fn}^{\mathrm{fit}}(\mathrm{Cs}), \rm sin^2 \theta_W\right)+\left(\frac{R_{\rm fn}^{\mathrm{fit}}(\mathrm{Cs})-R_{\rm fn}^{\mathrm{exp}}(\mathrm{Cs})}{\sigma ^{\mathrm{exp}}}\right)^{2},
\end{equation}
where $\sigma ^{\mathrm{exp}}$ is the uncertainty of $R^{\rm exp}_{\rm fn}(\mathrm{Cs})$.
The $\chi ^{2}_{\mathrm{CsI}}$ was constructed (see Ref.~\cite{Corona23} for details) and used for the free two-dimensional fit as mentioned above. 
The used $R^{\rm exp}_{\rm fn}(\mathrm{Cs})$ of 4.936(210) fm is obtained by the presently deduced point-neutron radius.
Now, for the first time we fix the range of neutron radii in the second term of Eq.~(6).
As a consequence, sin$^2 \theta_W$ of 0.227(28) is independently extracted from the best fit to the CE$\nu$NS-CsI data~\cite{Akimov221} as shown with the black symbol in Fig.~\ref{fig4}. 
The black curves represent the distributions of the sin$^2 \theta_W$ values obtained from fitting procedures for different CL.  
Our determined value sin$^2 \theta_W=0.227(28)$ agrees within error bars with the SM prediction of 0.23857(5) at low momentum transfer~\cite{Zyla20}. 
Compared to the sin$^2 \theta_W$ value of 0.309$^{+0.078}_{-0.063}$ determined by the free independent 2D fit, our analysis well constrains the sin$^2 \theta_W$ value and improves its precision by a factor of about 2.5.
However, the present precision of about 10$\%$ for our sin$^2 \theta_W$ value is still too large to effectively probe new physics.
This large uncertainty is mainly caused by the CE$\nu$NS-CsI data~\cite{Akimov221}, see the red symbol and blue area in the inset of Fig.~\ref{fig3}.
Therefore, the key neutron radius value provided in this work would play an important role in searching new physics related to the $\nu$ interactions via the high precision CE$\nu$NS-CsI data in the near future.

In conclusion, the proton elastic scattering off $^{133}$Cs at 199.4 MeV/u was investigated in inverse kinematics at HIRFL-CSR.
Combined with the proton distribution radius, a point-neutron radius of 4.86(21) fm for $^{133}$Cs was extracted.
For the first time we fix the range of neutron radii of $^{133}$Cs in the CE$\nu$NS-CsI data analysis, and consequently the weak mixing angle was extracted independently with higher accuracy than the ones proposed by the previous independent approaches.
This work provides a key neutron radius data and constitutes a new approach for the interpretation of high precision CE$\nu$NS-CsI data in the near future.


\section*{Acknowledgements}

We thank the staffs in the accelerator division for providing a stable beam. This work is supported in part by the National Key R\&D Program of China (Grant No. 2023YFA1606401),
by the NSFC (Grants No.~12375115, No.~12022504, No.~12121005, No.~12075255, and No.~12275186),
and by the CAS Open Research Project of large research infrastructures.


\begin{thebibliography}{00}

\bibitem{Suekane21}F. Suekane, Quantum Oscillations, Lecture Notes in Physics {\bf 985}, Springer Nature Switzerland AG 2021.
\bibitem{Kumar13}K. S. Kumar, S. Mantry, W. J. Marciano, and P. A. Souder, Annu. Rev. Nucl. Part. Sci. {\bf 63}, 237 (2013).
\bibitem{ALEPH06}The ALEPH Collaboration, The DELPHI Collaboration, The L3 Collaboration, The OPAL Collaboration, The SLD Collaboration, The LEP Electroweak Working Group, The SLD Electroweak, and Heavy Flavour Groups, Phys. Rep. {\bf 427}, 257 (2006).
\bibitem{CMS}The CMS Collaboration, Measurement of the Drell-Yan forward-backward asymmetry and of the effective leptonic weak mixing angle using proton-proton collisions at $\sqrt{s}$= 13 TeV, Report number: CMS PAS SMP-22-010 (2024).
\bibitem{Dzuba12}V. A. Dzuba, J. C. Berengut, V. V. Flambaum, and B. Roberts, Phys. Rev. Lett. {\bf 109}, 203003 (2012).
\bibitem{Wood97}C. S. Wood, S. C. Bennett, D. Cho, B. P. Masterson, J. L. Roberts, C. E. Tanner, and C. E. Wieman, Science {\bf 275}, 1759 (1997).
\bibitem{Cadeddu23}M. Cadeddu, F. Dordei, and C. Giunti, Europhys. Lett. {\bf 143}, 34001 (2023).
\bibitem{Sahoo21}B. K. Sahoo, B. P. Das, and H. Spiesberger,  Phys. Rev. D {\bf 103}, L111303 (2021).
\bibitem{Pollock99}S. J. Pollock and M. C. Welliver, Phys. Lett. B {\bf 464}, 177 (1999).
\bibitem{Akimov17}D. Akimov, J. B. Albert, P. An, C. Awe, P. S. Barbeau, B. Becker, V. Belov, A. Brown, A. Bolozdynya, B. Cabrera-Palmer \emph{et al.}, Science {\bf 357}, 1123 (2017).
\bibitem{Akimov221}D. Akimov, P. An, C. Awe, P. S. Barbeau, B. Becker, V. Belov, I. Bernardi, M. A. Blackston, C. Bock, A. Bolozdynya \emph{et al.}, Phys. Rev. Lett. {\bf 129}, 081801 (2022).
\bibitem{Akimov23}D. Akimov, P. An, C. Awe, P. S. Barbeau, B. Becker, V. Belov, I. Bernardi, M. A. Blackston, C. Bock, A. Bolozdynya \emph{et al.}, Phys. Rev. Lett. {\bf 130}, 051803 (2023).
\bibitem{Cadeddu2021jh}M. Cadeddu, N. Cargioli, F. Dordei, C. Giunti, Y. F. Li, E. Picciau, and Y. Y. Zhang, J. High Energy Phys. {\bf 01}, 116  (2021).
\bibitem{Barbeau23}P. S. Barbeau, Yu. Efremenko, and K. Scholberg, Annu. Rev. Nucl. Part. Sci. {\bf 73}, 41 (2023).
\bibitem{Huang19}X. -R. Huang and L. -W. Chen, Phys. Rev. D {\bf 100}, 071301(R) (2019).
\bibitem{Sierra19}D. Aristizabal Sierra, J. Liao, and D. Marfatia, J. High Energy Phys. {\bf 06}, 141 (2019) .
\bibitem{Papoulias20}D. K. Papoulias, Phys. Rev. D {\bf 102}, 113004 (2020).
\bibitem{Khan19}A. N. Khan and W. Rodejohann, Phys. Rev. D {\bf 100}, 113003 (2019).
\bibitem{Cadeddu20}M. Cadeddu, F. Dordei, C. Giunti, Y. F. Li, E. Picciau, and Y. Y. Zhang, Phys. Rev. D {\bf 102}, 015030 (2020).
\bibitem{Cadeddu18}M. Cadeddu, C. Giunti, Y. F. Li, and Y. Y. Zhang, Phys. Rev. Lett. {\bf 120}, 072501 (2018).
\bibitem{Cadeddu201}M. Cadeddu, F. Dordei, C. Giunti, Y. F. Li, and Y. Y. Zhang, Phys. Rev. D {\bf 101}, 033004 (2020).
\bibitem{Cadeddu21}M. Cadeddu, N. Cargioli, F. Dordei, C. Giunti, Y. F. Li, E. Picciau, C. A. Ternes, and Y. Y. Zhang, Phys. Rev. C {\bf 104}, 065502 (2021).
\bibitem{Coloma20}P. Coloma, I. Esteban, M. C. Gonzalez-Garcia, and J. Men\'endez, J. High Energy Phys. {\bf 08}, 030 (2020).
\bibitem{Cadeddu19}M. Cadeddu and F. Dordei, Phys. Rev. D {\bf 99}, 033010 (2019).
\bibitem{Romeri23}V. De Romeri, O. G. Miranda, D. K. Papoulias, G. Sanchez Garcia, M. T\'ortola, and J. W. F. Valle, J. High Energy Phys. {\bf 04}, 035 (2023).
\bibitem{Papoulias20_1}D. K. Papoulias, T. S. Kosmas, R. Sahu, V. K. B. Kota, and M. Hota, Phys. Lett. B {\bf 800}, 135133 (2020).
\bibitem{Corona23}M. Atzori. Corona, M. Cadeddu, N. Cargioli, F. Dordei, C. Giunti, and G. Masia, Eur. Phys. J. C {\bf 83}, 683 (2023).
\bibitem{Xia02}J. W. Xia, W. L. Zhan, B. W. Wei, Y. J. Yuan, M. T. Song, W. Z. Zhang, X. D. Yang, P. Yuan, D. Q. Gao, H. W. Zhao \emph{et al.}, Nucl. Instrum. Methods Phys. Res. A {\bf 488}, 11 (2002).
\bibitem{Liu20}X. Liu, P. Egelhof, O. Kiselev, and M. Mutterer, Phys. Lett. B {\bf 809}, 135776 (2020).
\bibitem{Zhang23}J. T. Zhang, P. Ma, Y. Huang, X. L. Tu, P. Sarriguren, Z. P. Li, Y. Kuang, W. Horiuchi, T. Inakura, L. Xayavong \emph{et al.}, Phys. Rev. C {\bf 108}, 014614 (2023).
\bibitem{Mao16}L. J. Mao, H. Zhao, X. D. Yang, J. Li, J. C. Yang, Y. J. Yuan, V. V. Parkhomchuk, V. B. Reva, X. M. Ma, T. L. Yan \emph{et al.}, Nucl. Instrum. Methods Phys. Res. A {\bf 808}, 29 (2016).
\bibitem{Shao13}C. J. Shao, R. C. Lu, X. H. Cai, D. Y. Yu, F. F. Ruan, Y. L. Xue, J. M. Zhang, D. K. Torpokov, and D. Nikolenko, Nucl. Instrum. Methods Phys. Res. B {\bf 317}, 617 (2013).
\bibitem{Yue19}K. Yue, J. T. Zhang, X. L. Tu, C. J. Shao, H. X. Li, P. Ma, B. Mei, X. C. Chen, Y. Y. Yang, X. Q. Liu \emph{et al.}, Phys. Rev. C {\bf 100}, 054609 (2019).
\bibitem{Zhang19}J. T. Zhang, K. Yue, H. X. Li, X. L. Tu, C. J. Shao, P. Ma, B. Mei, X. C. Chen, Y. Y. Yang, X. Q. Liu \emph{et al.}, Nucl. Instrum. Methods Phys. Res. A {\bf 948}, 162848 (2019).
\bibitem{Khazov11}Yu. Khazov, A. Rodionov, and F. G. Kondev, Nucl. Data Sheets {\bf 112}, 855 (2011).
\bibitem{McDaniels86}D. K. McDaniels, J. R. Tinsley, J. Lisantti, D. M. Drake, I. Bergqvist, L. W. Swenson, F. E. Bertrand, E. E. Gross, D. J. Horen, T. P. Sjoreen \emph{et al.}, Phys. Rev. C {\bf 33}, 1943 (1986).
\bibitem{Kailas84}S. Kailas, P. P. Singh, D. L. Friesel, C. C. Foster, P. Schwandt, and J. Wiggins, Phys. Rev. C {\bf 29}, 2075 (1984).
\bibitem{Thompson88}I. J. Thompson, Comput. Phys. Rep. {\bf 7}, 167 (1988); http://www.fresco.org.uk/.
\bibitem{Koning03}A. J. Koning and J. P. Delaroche, Nucl. Phys. A {\bf 713}, 231 (2003).
\bibitem{Alkhazov97}G. D. Alkhazov, M. N. Andronenko, A. V. Dobrovolsky, P. Egelhof, G. E. Gavrilov, H. Geissel, H. Irnich, A. V. Khanzadeev, G. A. Korolev, A. A. Lobodenko \emph{et al.}, Phys. Rev. Lett. {\bf 78}, 2313 (1997).
\bibitem{Korolev18}G. A. Korolev, A. V. Dobrovolsky, A. G. Inglessi, G. D. Alkhazov, P. Egelhof, A. Estrad\'e, I. Dillmann, F. Farinon, H. Geissel, S. Ilieva \emph{et al.}, Phys. Lett. B {\bf 780}, 200 (2018).
\bibitem{Schmid23}M. von Schmid, T. Aumann, S. Bagchi, S. B\"onig, M. Csatl\'os, I. Dillmann, C. Dimopoulou, P. Egelhof, V. Eremin, T. Furuno \emph{et al.}, Eur. Phys. J. A {\bf 59}, 83 (2023).
\bibitem{Huang232}Y. Huang, L. Xayavong, X. L. Tu, J. Geng, Z. P. Li, J. T. Zhang, and Z. H. Li, Phys. Lett. B {\bf 847}, 138293 (2023).
\bibitem{Alkhazov78}G. D. Alkhazov, S. L. Belostotsky, and A. A. Vorobyov, Phys. Rep. {\bf 42}, 89 (1978).
\bibitem{Kudryavtsev89}I. N. Kudryavtsev and A. P. Soznik, J. Phys. G: Nucl. Part. Phys. {\bf 15}, 1377 (1989).
\bibitem{Thiel19}M. Thiel, C. Sfienti, J. Piekarewicz, C. J. Horowitz, and M. Vanderhaeghen, J. Phys. G: Nucl. Part. Phys. {\bf 46}, 093003 (2019).
\bibitem{Terashima08}S. Terashima, H. Sakaguchi, H. Takeda, T. Ishikawa, M. Itoh, T. Kawabata, T. Murakami, M. Uchida, Y. Yasuda, M. Yosoi \emph{et al.}, Phys. Rev. C {\bf 77}, 024317 (2008).
\bibitem{Huang23}Y. Huang, J. T. Zhang, Y. Kuang, J. Geng, X. L. Tu, K. Yue, W. H. Long, and Z. P. Li, Eur. Phys. J. A {\bf 59}, 4 (2023).
\bibitem{Alkhazov02}G. D. Alkhazov, A. V. Dobrovolsky, P. Egelhof, H. Geissel, H. Irnich, A. V. Khanzadeev, G. A. Korolev, A. A. Lobodenko, G. M\"unzenberg, M. Mutterer \emph{et al.}, Nucl. Phys. A {\bf 712}, 269 (2002).
\bibitem{Huang231}Y. Huang, X. Y. Wu, X. L. Tu, Z. P. Li, Y. Kuang, J. T. Zhang, and Z. H. Li,  Phys. Rev. C {\bf 108}, 054610 (2023).
\bibitem{Glauber59}R. J. Glauber, in \textit{Lectures in Theoretical Physics}, edited by W. E. Brittin and L. G. Dunham (Interscience, New York, 1959), Vol. 1, p. 315.
\bibitem{Tanaka20}M. Tanaka, M. Takechi, A. Homma, M. Fukuda, D. Nishimura, T. Suzuki, Y. Tanaka, T. Moriguchi, D. S. Ahn, A. Aimaganbetov \emph{et al.}, Phys. Rev. Lett. {\bf 124}, 102501 (2020).
\bibitem{Suzuki95}T. Suzuki, H. Geissel, O. Bochkarev, L. Chulkov, M. Golovkov, D. Hirata, H. Irnich, Z. Janas, H. Keller, T. Kobayashi \emph{et al.}, Phys. Rev. Lett. {\bf 75}, 3241 (1995).
\bibitem{Angeli13}I. Angeli and K. P. Marinova, At. Data Nucl. Data Tables {\bf 99}, 69 (2013).
\bibitem{2021Cadeddu}M. Cadeddu, N. Cargioli, F. Dordei, C. Giunti, and E. Picciau, Phys. Rev. D {\bf 104}, L011701 (2021).
\bibitem{Adhikari21}D. Adhikari, H. Albataineh, D. Androic, K. Aniol, D. S. Armstrong, T. Averett, C. Ayerbe Gayoso, S. Barcus, V. Bellini, R. S. Beminiwattha \emph{et al.}, Phys. Rev. Lett. {\bf 126}, 172502 (2021).
\bibitem{Adhikari22}D. Adhikari, H. Albataineh, D. Androic, K. A. Aniol, D. S. Armstrong, T. Averett, C. Ayerbe Gayoso, S. K. Barcus, V. Bellini, R. S. Beminiwattha \emph{et al.}, Phys. Rev. Lett. {\bf 129}, 042501 (2022).
\bibitem{Trzcinska01}A. Trzci\'nska, J. Jastrz\c{e}bski, P. Lubi\'nski, F. J. Hartmann, R. Schmidt, T. von Egidy, and B. K{\l}os, Phys. Rev. Lett. {\bf 87}, 082501 (2001).
\bibitem{Derevianko01}A. Derevianko, Phys. Rev. A {\bf 65}, 012106 (2001).
\bibitem{Zyla20}P. A. Zyla, R. M. Barnett, J. Beringer, O. Dahl, D. A. Dwyer, D. E. Groom, C. J. Lin, K. S. Lugovsky, E. Pianori, D. J. Robinson \emph{et al.} (Particle Data Group), Prog. Theor. Exp. Phys. {\bf 2020}, 083C01 (2020).


\end{thebibliography}
\end{document}